%% file: main_ccta.tex
\documentclass[letterpaper,conference]{ieeeconf}
\IEEEoverridecommandlockouts                              
\usepackage[utf8]{inputenc}
\usepackage{textcomp}

\usepackage{graphicx}
\usepackage{amsmath}
\usepackage[version=4]{mhchem}
\usepackage{siunitx}
\usepackage{graphicx}
\usepackage{longtable,tabularx}
\usepackage[backend=biber,style=ieee ,sorting=none,maxbibnames=99,giveninits, doi=true]{biblatex}
\AtBeginBibliography{\small}
\input{PackagesCommands}
\usepackage{comment}

\title{Adaptive Control of multirotor in Continuous Time}

\title{Single-shot Learning of a multirotor Controller}

\title{Transfer Learning of a Multirotor Control System}

\title{Learning a Multirotor Control System via Transfer Learning}

\title{Multirotor Controller Training using Transfer Learning}

\title{Multirotor Controller Single-shot Training in a Transfer Learning Framework \\ based on  Retrospective Cost Optimization}

\title{Sim-to-Real Multirotor Controller Learning \\ based on Single-shot Retrospective Cost Optimization}

\title{Sim-to-Real Multirotor Controller Single-shot Learning }

\title{Retrospective Cost Optimization for Multirotor Controller Autotuning: Single-Shot Learning in Simulation for Physical Implementation}

\title{Retrospective Cost Optimization for Multirotor Controller Autotuning: Single-Shot Learning in Simulation for Physical Implementation}

\title{A Single Shot Learning Approach for Multirotor Controller \\ with Experimental Validation}

\title{Single-Shot Learning of Multirotor Controller Gains: \\ A Data-Driven Approach with Experimental Validation}

\author{
Mohammad Mirtaba, 
Parham Oveissi, 
Juan Augusto Paredes Salazar,
and
Ankit Goel
\thanks{Mohammad Mirtaba, Parham Oveissi, Juan Augusto Paredes Salazar, and Ankit Goel are with the Department of Mechanical Engineering, University of Maryland, Baltimore County, 1000 Hilltop Circle, Baltimore, MD, 21250. {\tt\small \{mmirtab1, parhamo1 japarede, ankgoel\}@umbc.edu }}%
}

\begin{document}

\maketitle


\begin{abstract}

This paper demonstrates the single-shot learning capabilities of retrospective cost optimization-based data-driven control applied to learning multirotor controller gains for trajectory tracking.
In particular, the proposed control approach is first used within a simple multirotor simulation environment to learn appropriate multirotor controller gains to follow a trajectory.
Then, the gains resulting from a single simulation run are used in a more complex multirotor simulation environment based on Simulink for performance verification.
Finally, the resulting gains are implemented in a physical quadrotor and the results for waypoint and trajectory tracking are reported in this paper.
The proposed control approach is the continuous-time version of the widely used discrete-time retrospective control adaptive control algorithm, which is simpler to implement within continuous-time simulation environments and whose performance does not depend on appropriate sampling time choice.

%

\end{abstract}

\section{Introduction}



Multirotor unmanned aerial vehicles are being increasingly used in various sectors due to their simple structure and vertical take-off and landing capabilities.
Applications now include a wide variety, such as agricultural applications \cite{app1,paredes2017} and search and rescue missions \cite{app2}, \cite{soleimani2021multiagent}.
Despite multirotors' successes, commercial flight controllers and autopilots are only well-tuned for well-established airframes. 
Due to underactuated nonlinear dynamics, parasitic vibrational effects \cite{mirtaba2023design}, and fluid-structure interactions, which are extremely difficult to model, model-based control design for multirotors is impractical. 
The uncertainty and time-varying nature of the physical parameters make this approach increasingly difficult. 
Furthermore, changes in the payload or physical modifications due to component upgrades degrade the nominal controller performance.

Learning-based, data-driven, and adaptive control techniques are thus ideally suited to design control systems for multirotor systems since such techniques can compensate for unknown, unmodeled, and uncertain dynamics. 
Several learning-based techniques have been explored for controller design. 
In particular, reinforcement learning \cite{rlutah, rlattitude, rlsigwart, rldavide, fullrl1, fullrl2, autotunerl, rlgeo} and particle swarm optimization \cite{mac2016,noordin2017,fessi2019,liu2020,gomez2020,guo2022,portillo2023} have been explored to design multirotor controllers. 
%
%
However, these approaches require several simulation iterations to generate sufficiently rich data to \textit{learn} the controller.
A nature-inspired evolutionary optimization algorithm is used to tune the parameters of the PID controllers in \cite{nature-auto}, where the squared error integral is used as a fitness function and minimized over generations in the context of evolutionary optimization.
However, as the authors note, the position and attitude controllers are separately tuned and require one of them to be well tuned to tune the other. 
Auto differentiation for gradient-based optimization is explored in \cite{difftune}, although this approach requires a sufficiently realistic model of the system. 
%

%
The present work uses
retrospective cost adaptive control (RCAC), which is a discrete-time output feedback adaptive control technique that can be applied to stabilization, command-following, and disturbance rejection problems.
The RCAC algorithm, its connection to linear quadratic control, and its extension to adaptive PID control are described in detail in \cite{rahmanCSM2017, rezaPID}.
%
%
%
In contrast to L1 adaptive control \cite{lu2019,hussien2021,wu2022} and adaptive MPC \cite{didier2021,hanover2021,zhang2024} schemes,
RCAC does not require a prior stabilizing controller and only requires minimal information about the system;
furthermore, neither control scheme provides a structure for controller gain learning, whereas RCAC can accommodate several linear control schemes for the learning procedure.
The discrete-time RCAC technique has been successfully applied to learn multirotor controllers in a single experiment \cite{goel2021experimental,parham}.
However, the discrete-time control design is vulnerable to sampling time choice, and the learned controller does not have a direct interpretation in terms of stability and transient performance.

A key contribution of this work is the development and application of a continuous-time retrospective cost adaptive control (CT-RCAC) for learning the multirotor controller gains. 
The CT-RCAC algorithm proves especially useful for applications typically simulated within a continuous time framework, such as CFD models. 
In this work, CT-RCAC is directly integrated with the Simulink framework that propagates continuous-time equations of motion of a multirotor. 
Specifically, CT-RCAC is used to learn the controller gains for a 12-degree-of-freedom (12dof) ideal multirotor model in Simulink. 
The learned controller is then validated against a realistic multirotor model in Simulink and the Holybro X500 V2 multirotor in physical flight tests. 
In the terminology of transfer learning \cite{transferlearning_survey}, the 12dof ideal multirotor model in Simulink serves the \textit{source environment}, and the realistic model in Simulink and the Holybro X500 V2 multirotor serve as the \textit{target environment}, as shown in Figure \ref{fig:learning_loop}.

\begin{figure}[H]
    \vspace{-0.5em}
    \centering
    \resizebox{\columnwidth}{!}
    {
    \begin{tikzpicture}[>={stealth'}, line width = 0.25mm]
    
        \draw [fill=green!20, rounded corners] (-1.5,-1.8) rectangle (4.8,1.5) node[below,xshift = -9em, below]
        {\textbf{Learning}};

        \draw [fill=red!20, rounded corners] (6.4,-1) rectangle +(3.5,2.4) node[below,xshift = -5em, below]
        {\textbf{Validation}};
        
        \node [smallblock, minimum height=3em] at (0,0) (Pos_Cont) {$\begin{array}{c} \mbox{Source} \\ \mbox{Environment} \\ \mbox{\scriptsize (12dof model)} \end{array}$};

        \node [smallblock, minimum height=3em, minimum width = 5em,  right = 3em of Pos_Cont] (learning) {$\begin{array}{c} \mbox{Data-driven} \\ \mbox{Learning} \\ \mbox{\scriptsize (CT-RCAC)}\end{array}$};

        \node [smallblock, minimum height=3em, minimum width = 5em,  right = 6em of learning] (target) {$\begin{array}{c} \mbox{Target} \\ \mbox{Environment}  \\ \mbox{\scriptsize (Realistic model, X500)} \end{array}$};
        
        %
                
        \draw [blue,->] (Pos_Cont) -- node  [xshift = 0em, above] {Data}  (learning.180);

        \draw [blue,->] (learning.-50) |- node[below,xshift = -6.5em, above]
        {Hyperparameter Tuning}
        ([xshift = -1.5em, yshift = -2 em]Pos_Cont.south) -- ([xshift = -1.5em]Pos_Cont.south);
        \draw [blue,->] (learning.0) -- node[below,xshift = 0em, above]
        {$\begin{array}{c} \mbox{Learned} \\ \mbox{Gains}  \end{array}$} (target);
    \end{tikzpicture}
    }
    \vspace{-0.75em}
    \caption{Learning and validating multirotor controller gains. }
    \vspace{-1em}
    \label{fig:learning_loop}
\end{figure}

We emphasize that all computational models represented by continuous-time differential equations are propagated discretely in digital computers. 
However, specialized simulation frameworks like Simulink utilize high-order accurate integration schemes to reduce computational demands without sacrificing solution accuracy. 
Therefore, solutions calculated by \textit{continuous-time solvers} are generally more accurate than the first-order Euler integration technique typically used to discretize the system model for discrete controller design.

The paper is organized as follows. 
Section \ref{sec:QuadDynCon} briefly reviews the multirotor dynamics and the control system architecture used in this work, 
Section \ref{sec:CT-RCAC} presents the continuous-time retrospective cost adaptive control algorithm, 
and 
Section \ref{sec:transfer_learning} describes the controller gain learning framework applied to a multirotor and shows results from simulations and physical experiments.
%
%
The paper concludes with a summary in Section \ref{sec:conclusions}.

\section{Multirotor Dynamics and Control}
\label{sec:QuadDynCon}

This section briefly overviews the 12dof multirotor dynamics and the traditional inner-outer control architecture. The ideal 12dof multirotor dynamics will act as the environment for learning controller gains. 
A detailed derivation of the multirotor dynamics is given in \cite{goel2020adaptive,goel2021experimental}.

\subsection{Multirotor Dynamics}
\label{sec:QuadDyn}
A multirotor can be modeled as a rigid body with a force applied along a body-fixed axis and a torque applied to the rigid body. 
The translational dynamics of a multirotor is
\begin{gather}
    m \ddot r = mg e_3 + \SO e_3 f,
    \label{eq:rddot}
\end{gather}
where 
$m$ is the mass of the multirotor, 
$r \in \BBR^3$ is the position of multirotor relative to a fixed point in an inertial frame, 
$g$ is the acceleration due to gravity, 
$e_3 \in \BBR^3$ is the third column of the $3\times 3$ identity matrix, 
$f \in \BBR$ is the force applied along a body-fixed axis, and 
the orthonormal matrix 
$\SO \in \BBR^{3\times 3}$ parameterizes the multirotor's attitude relative to an inertial frame.

The rotational dynamics of a multirotor is
\begin{gather}
    \dot \SO = -\omega^\times \SO ,
    \label{eq:poissons-eqn}
    \\
    J \dot \omega + \omega \times J \omega = \tau, 
    \label{eq:eulers-eqn}
\end{gather}
where $\omega \in \BBR^3$ is the angular velocity vector of the multirotor relative to an inertial frame resolved in the body-fixed frame, 
$J \in \BBR^{3 \times 3}$ is the moment of inertia matrix, and 
$\tau \in \BBR^3$ is the torque applied to the multirotor in the body-fixed frame. 

Note that \eqref{eq:rddot} models the translational motion of the multirotor, whereas \eqref{eq:poissons-eqn}, \eqref{eq:eulers-eqn} model the rotational motion of the multirotor. 
This separation of translational and rotational motions facilitates the implementation of outer and inner loop architectures for trajectory tracking.
%
%
However, note that the translational motion and the rotational motion of the multirotor are coupled via the attitude $\SO$ in \eqref{eq:rddot}, and thus the outer and inner loop design cannot be entirely decoupled.
In fact, it is precisely the coupling due to the $\SO$ matrix which allows stabilization of the underactuated translational dynamics.

\subsection{Multirotor Control}
\label{sec:autopilot}
\begin{figure}[h!]
    \vspace{-1em}   
    \centering
    \resizebox{\columnwidth}{!}
    {
    \begin{tikzpicture}[>={stealth'}, line width = 0.25mm, text centered]
    
        \node [smallblock, minimum height=3em, text width=1.6cm] (Mission) { Mission Planner};
        \node [smallblock, minimum height=3em, right = 7em of Mission, text width=1.6cm] (Pos_Cont) { Position Controller};

        \node [right = 3 em of Pos_Cont] (midpoint) {};
        
        \node [smallblock, minimum height=3em, below = 1.5em of midpoint,text width=1.6cm] (Att_Cont) {Attitude Controller};
        \node [smallblock, minimum height=3em, minimum width = 5.5em,  right = 7em of midpoint] (multirotor) { Multirotor};
        
        \draw [blue,->] (Mission) -- node[above, xshift = -0.05 em, text width=1.6cm]{Position reference} (Pos_Cont);
        \draw[red,->] (Mission.-90) |- +(0,-.5) 
        node[xshift = 6 em, yshift = 0.25 em]{Yaw reference}
        |- (Att_Cont.180);
        %
        

        \draw [red,->] (Pos_Cont.15) -| (Att_Cont.90);
        
        \draw [blue,->] (Pos_Cont.15) -- node  [xshift = 0em, above] {Force} (multirotor.165);
        
        \draw [red,->] (Att_Cont.0) -- +(0.1,0) |-  node [below, xshift = 2 em]{Moment}(multirotor.195);
        \draw [blue,->] (multirotor.15) -- +(1,0) |- node[below,xshift = -10em, above]{Position and Velocity Measurement} ([yshift = -7.5 em]Pos_Cont.south) -- (Pos_Cont.south);
        \draw [red,->] (multirotor.345) -- +(0.5,0) |- node[above,xshift = -7em]{Attitude and Rate Measurement} ([yshift = -2 em]Att_Cont.south) -- (Att_Cont.south);

    \end{tikzpicture}
    }
    \vspace{-0.5em}    
    \caption{Cascaded loop control architecture for a multirotor. }
    \vspace{-1em}   
    \label{fig:autopilot_nested_loop}
\end{figure}
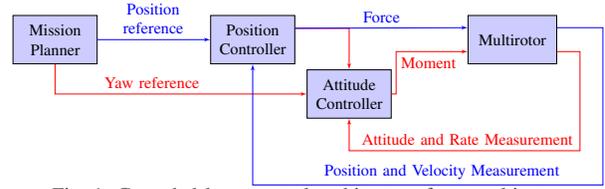

This work considers a cascaded control system, as shown in Figure \ref{fig:autopilot_nested_loop}.
The cascaded loop architecture is motivated by the time separation principle. 
The outer loop is designed to track the position references, whereas the inner loop is designed to track the attitude references.
The cascaded loop architecture assumes that the inner loop dynamics are significantly faster than the outer loop dynamics. 
In this case, the three-dimensional force vector required to move the multirotor in a desired direction can be realized by adjusting the attitude of the multirotor so that the axis of the multirotor is along the direction of the force vector and modulating the total force exerted by the propellers.

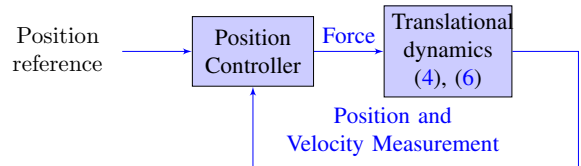
\begin{figure}[h!]
    \vspace{-0.75em}  
    \centering
    \resizebox{0.8\columnwidth}{!}
    {
    \begin{tikzpicture}[>={stealth'}, line width = 0.25mm]
    
        \node (Mission) 
        {
            $\begin{array}{c}
                \rm Position \\
                \rm reference 
            \end{array}$            
        };
        \node [smallblock, minimum height=3em, right = 1em of Mission] (Pos_Cont) {$\begin{array}{c} \mbox{Position} \\ \mbox{Controller} \end{array}$};

        \node [smallblock, minimum height=3em, minimum width = 7em,  right = 4em of Pos_Cont] (multirotor) {$\begin{array}{c} \mbox{Translational} \\ \mbox{Dynamics} \\ \eqref{eq:r1_outer_loop_eqn}, \eqref{eq:r3_outer_loop_eqn} \end{array}$};
        
        \draw [blue, ->] ([xshift = -0.4em]Mission.east) -- (Pos_Cont);

        \draw [blue,->] (Pos_Cont) -- node  [xshift = 0em, above] {Force}  (multirotor.180);

        \draw [blue,->] (multirotor.0) -- +(0.5,0) |- node[below,xshift = -8em, above]
        {
            $\begin{array}{c}
                \text{Position and} \\
                \text{Velocity Measurement}
            \end{array}$            
        }
        ([yshift = -4 em]Pos_Cont.south) -- (Pos_Cont.south);

    \end{tikzpicture}
    }
    \vspace{0.5em}
    \caption{Outer loop to track position reference. }
    \vspace{-1em}   
    \label{fig:outer_loop}
\end{figure}

Since the inner loop, which regulates the rotational motion, is assumed to be significantly faster than the outer loop, the translational motion of the multirotor satisfies
\begin{align}
    m \ddot r_1 &= f_{r_1},
    \label{eq:r1_outer_loop_eqn}
    \\
    m \ddot r_2 &= f_{r_2},
    \label{eq:r2_outer_loop_eqn}
    \\
    m \ddot r_3 &= mg +  f_{r_3},
    \label{eq:r3_outer_loop_eqn}
\end{align}
where $r_1, r_2,$ and $r_3$ are the components of $r,$
$f_{r_1}$ and $f_{r_2}$ are the horizontal forces and $f_{r_3}$ is the vertical force on the multirotor. 
Note that \eqref{eq:r1_outer_loop_eqn}-\eqref{eq:r3_outer_loop_eqn} are linear and decoupled, where the \textit{effective} forces 
%
%
$f_{r_1}, f_{r_2}$ and $f_{r_3}$ are assumed to be realizable due to the faster rotational dynamics.
The objective of the outer loop is thus to compute the force input to track a desired position reference, as shown in Figure \ref{fig:outer_loop}.
The position controller considered in this work consists of three decoupled controllers, each tasked to follow the corresponding position command.

\begin{figure}[h!]
    \vspace{-1.5em}
    \centering
    \resizebox{0.8\columnwidth}{!}
    {
    \begin{tikzpicture}[>={stealth'}, line width = 0.25mm]
    
        \node (Mission) 
        {
            $\begin{array}{c}
                \rm Attitude \\
                \rm reference 
            \end{array}$            
        };
        \node [smallblock, minimum height=3em, right = 1em of Mission] (Pos_Cont) {$\begin{array}{c} \mbox{Attitude} \\  \mbox{Controller} \end{array}$};

        \node [smallblock, minimum height=3em, minimum width = 5.5em,  right = 4.5em of Pos_Cont] (multirotor) {$\begin{array}{c} \mbox{Rotational} \\ \mbox{Dynamics} \\ \eqref{eq:eulers-eqn}, \eqref{eq:EulerDots}\end{array}$};
        
        \draw [red,->] ([xshift = -0.5em]Mission.east) -- (Pos_Cont);

        \draw [red,->] (Pos_Cont) -- node  [xshift = 0em, above] {Torque}  (multirotor.180);

        \draw [red,->] (multirotor.0) -- +(0.5,0) |- node[below,xshift = -8em, above]
        {
            $\begin{array}{c}
                \text{Attitude and} \\
                \text{Rate Measurement}
            \end{array}$            
        }
        ([yshift = -4 em]Pos_Cont.south) -- (Pos_Cont.south);

    \end{tikzpicture}
    }
    \vspace{0.5em}
    \caption{Inner loop to track attitude reference. }
    \vspace{-1em}
    \label{fig:inner_loop}
\end{figure}
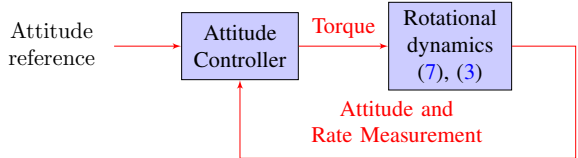

The inner loop computes the moment required to realize the force vector desired by the outer loop by orienting the multirotor so that its force axis aligns with the desired force vector required by the outer loop. 
The computation of the desired attitude is described in detail in \cite{goel2021experimental}.
%
In terms of the 3-2-1 Euler angles, representing the yaw, roll, and pitch,  \eqref{eq:poissons-eqn} is
\begin{align}
    \dot \Theta = S(\Theta) \omega, 
    \label{eq:EulerDots}
\end{align}
where $\Theta \in \BBR^3$ contains the three Euler angles. 
Note that the rotational motion of the multirotor is governed by the nonlinear coupled dynamics \eqref{eq:eulers-eqn}, \eqref{eq:EulerDots}.
However, it is well known that for small Euler angles, the rotational dynamics consists of three decoupled double integrators similar to \eqref{eq:r1_outer_loop_eqn}-\eqref{eq:r3_outer_loop_eqn}.
Figure \ref{fig:inner_loop} shows the inner loop to track the desired attitude reference.
The attitude controller considered in this work consists of three decoupled controllers, each tasked to follow the corresponding angle command. 

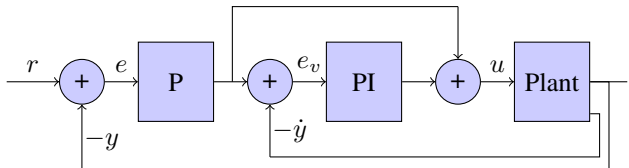
\begin{figure}[h]
\vspace{-0.5em}
    \centering
    \resizebox{0.9\columnwidth}{!}{%
    \begin{tikzpicture}[>={stealth'}, line width = 0.25mm]
        \node [input, name=input] {};
        \node [sum, right = 2 em of input, inner sep = 0.1em, minimum height=0.1em, name=sum1] {+};
        \node [block, right = 2 em of sum1, name=PCONTROL, minimum width=2.5em, minimum height=2.5em] {P};
        \node [output, right = 0.6em of PCONTROL,node distance = 0.75cm] (outputp) {};
        \node [sum, right = 1.5em of PCONTROL,inner sep = 0.1em, minimum height=0.1em,name=sum2] {+};
        \node [block, right = 2 em of sum2, name=PICONTROL, minimum width=2.5em, minimum height=2.5em] {PI};
        \node [sum, right = 1 em of PICONTROL, name=sum3, inner sep = 0.1em, minimum height=0.1em] {+};
        \node [block, right = 2em of sum3, name=PLANT, minimum width=2.5em, minimum height=4.5em] {Plant};
        \node [output, right = 3.5em of PLANT] (output1) {};
        \node [input, below = 3.5em of sum1, name=undersum1] {};
        \node [input, below = 2.5em of sum2, name=undersum2] {};
        \node [input, above = 1.75em of sum3] (abovesum3) {};
    
        \draw [->] (input) -- node[above] {$r$} (sum1);
        \draw [->] (sum1) -- node[above] {$e$} (PCONTROL);
        \draw [->] (PCONTROL) -- (sum2);
        \draw [->] (sum2) -- node[above] {$e_v$} (PICONTROL);
        \draw [->] (PICONTROL) -- (sum3);
        \draw [->] (sum3) -- node[above] {$u$} (PLANT);
        \draw [->] (PLANT) -- node [xshift = -0.1em, above, near start] {$y$} (output1);
        \draw [-] (PLANT.0) -| +(0.75,-0.5) |- (undersum1);
        \draw [-] (PLANT.310) -| node [above, near start] {$\dot{y}$} +(0.50,-0.5) |- (undersum2);
        \draw [->] (undersum1) -- node [xshift = 0.5em, yshift = 1em] {$-$}(sum1);
        \draw [->] (undersum2) -- node [xshift = 0.5em, yshift = 0.5em] {$-$} (sum2);
        \draw [-] (outputp) |- (abovesum3);
        \draw [->] (abovesum3) -| (sum3);
    \end{tikzpicture}
    }
    \vspace{0.75em}
    \caption{Modified cascaded P-PI with control architecture for command following.}
    \label{fig:cascaded-ppi}
    \vspace{-1.5em}
\end{figure}

To follow the position and angle references in the outer and the inner loop, we consider a modified cascaded P-PI architecture shown in Figure \ref{fig:cascaded-ppi}. 
As shown in Figure \ref{fig:cascaded-ppi}, the control $u$ is given by
\begin{align}
    u = k_{\rmp,1} e + k_{\rmp,2} e_v + k_{\rmi} \int e_v,
    \label{eq:u_PID}
\end{align}
where
the tracking error $e \isdef r-y$, 
$k_{\rmp,1}$ and $k_{\rmp,2}$ are the proportional gains, $k_{\rmi}$ is the integral gain, and  
$e_v \isdef k_{\rmp,1} e - \dot y.$
Note that the controller \eqref{eq:u_PID} can be written as 
\begin{align}
    u(t) = \phi(t) \theta, 
    \label{eq:u_ppi}
\end{align}
where the data regressor $\phi(t) \isdef \matl  e & e_v & \int e_v \matr $ and the controller gain $\theta \isdef \matl k_{\rmp,1} &  k_{\rmp,2} & k_{\rmi}  \matr^\rmT. $
The controller gain $\theta$ is learned by retrospective cost optimize as described below. 

\section{Online Continuous Time Learning}
\label{sec:CT-RCAC}
This section describes the continuous-time retrospective cost adaptive control (CT-RCAC) used to optimize controller gains in this work.

Consider a dynamic system where $u(t) \in \BBR^{l_u}$ is the input, 
$y(t) \in \BBR^{l_y}$ is the measured output.
The objective is to design an adaptive output feedback control law using only the measured $y(t)$ and the input $u(t)$ without using the underlying dynamics.
%
%
Define the \textit{performance variable}
\begin{align}
    z(t) \isdef y(t) - r(t), 
    \label{eq:z_def}
\end{align}
where $r(t)$ is the exogenous reference signal. 
The objective of the adaptive output feedback control law is thus to ensure that $z(t) \to 0.$

Consider a linearly parameterized control law 
\begin{align}
    u(t) 
        =
            \Phi(t) \theta(t), 
    \label{eq:u_para}
\end{align}
where the regressor matrix $\Phi(t) \in \BBR^{l_u \times l_\theta}$ contains the measured data and 
the vector $\theta(t) \in \BBR^{l_\theta} $  contains the controller gains to be optimized. 
Various linear parameterizations of MIMO controllers are described in \cite{goel_2020_sparse_para}.
In this work, we use the SISO controller shown in \eqref{eq:u_ppi}.

Next, using \eqref{eq:z_def}, define the \textit{retrospective performance}
\begin{align}
    \hat z(t) 
        &\isdef
            z(t) + \Phi_\rmf(t) \hat \theta(t) -u_\rmf(t),
\end{align}
where $\hat \theta(t)$ is the controller gain to be optimized and  
the filtered regressor $\Phi_\rmf(t)$ and the filtered control $u_\rmf(t)$ are defined as
\begin{align}
    \Phi_\rmf(t) &\isdef G_\rmf(s) \left[ \Phi(t) \right] \in \BBR^{l_y \times l_\theta},
    \\
    u_\rmf(t) &\isdef G_\rmf(s) \left[ u(t) \right] \in \BBR^{l_y \times l_\theta},
\end{align}
where $G_\rmf(s)$ is a dynamic filter.

Next, define the \textit{retrospective cost} 
\begin{align}
    J(t, \hat \theta)
        &=
            \int_{0}^{t} 
                \hat z(\tau)^\rmT R_z \hat z(\tau) 
            \rmd \tau
            + \hat \theta^T R_\theta \hat \theta,
    \label{eq:J_RCAC}
\end{align}
where $R_z$
and $R_\theta$ are positive definite weighting matrices of appropriate dimensions.

\begin{proposition}
    \label{prop:RCAC_minimizer}
    Consider the cost function $J(t, \hat \theta)$ given by \eqref{eq:J_RCAC}.
    For all $t \ge 0, $ define the minimizer of $J(t, \hat \theta)$ by
    \begin{align}
        \theta^*(t)
            \isdef 
                \underset{\hat \theta \in \BBR^{l_\theta}}{\operatorname{argmin}} 
                J(t, \hat \theta).
    \end{align}
    Then, for all $t \ge 0, $ the minimizer is given by
    \begin{align}
        \theta^*(t)
            =
                P(t) b(t), 
    \end{align}
    where 
    \begin{align}
        \dot P(t) 
        &=
            -P
            \Phi_\rmf ^\rmT 
                    R_z
                    \Phi_\rmf  
            P
            , \\
        \dot b(t)
        &=
            2\Phi_\rmf^\rmT
                R_z 
                (z-u_\rmf),
    \end{align}
    and $P(0) = R_\theta^{-1}$ and $b(0) = 0.$
    
\end{proposition}
\begin{proof}
    Note that the cost function \eqref{eq:J_RCAC} can be written as
\begin{align}
    J(t, \hat \theta) 
        &=
            \hat \theta^T A(t) \hat \theta + \hat \theta^T b(t) + c(t),
    \label{eq:J_reformulated}
\end{align}
where
\begin{align}
    A(t) 
        &\isdef
            \int_0^{t} 
                    \Phi_\rmf ^\rmT 
                    R_z
                    \Phi_\rmf  
            \rmd \tau + R_\theta,
    \nn
    \\
    b(t) 
        &\isdef
            \int_0^{t}
            2\Phi_\rmf^\rmT
                R_z 
                (z-u_\rmf) \rmd \tau, 
    \nn \\
    c(t) 
        &\isdef
            \int_0^{t}
            (z-u_\rmf)^\rmT
                R_z 
                (z-u_\rmf)  
                \rmd \tau.
    \nn
\end{align}
Next, note that $A(t)$ and $b(t)$ satisfy
\begin{align}
    \dot A
        &=
            \Phi_\rmf ^\rmT 
                    R_z
                    \Phi_\rmf  
    \\
    \dot b
        &=
            2\Phi_\rmf^\rmT
                R_z 
                (z-u_\rmf) , \nn 
\end{align}
where $A(0) = R_\theta$ and $b(0) = 0.$


Next, define, for all $\ge 0,$ $P(t) \isdef A(t)^{-1}.$
Using the fact that $P(t) A(t) = I,$ it follows that
\begin{align}
    \dot P(t) = - P(t) \dot A(t) P(t), \nn
\end{align}
and thus 
\begin{align}
    \dot P 
        =
            -P
            \Phi_\rmf ^\rmT 
                    R_z
                    \Phi_\rmf  
            P. \nn
\end{align}
Finally, the minimizer of \eqref{eq:J_reformulated} is given by
\begin{align}
    \theta^{*}(t)
        =
            A(t)^{-1} b(t)
        =
            P(t) b(t), \nn
\end{align}
which completes the proof. 
\end{proof}

For all $t \geq 0,$ the control is given by
\begin{align}
    u(t) = \Phi(t) \theta^*(t).
\end{align}

\section{Learning Controller Gains}
\label{sec:transfer_learning}

\begin{figure}[h!]
    \vspace{-0.5em}
    \centering
    \includegraphics[width=0.5\columnwidth]{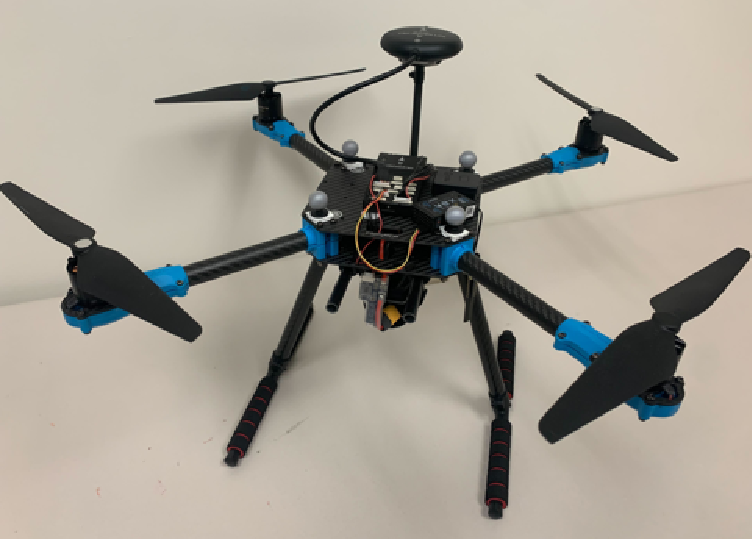}
    \vspace{0.5em}
    \caption{
    Holybro X500 V2 multirotor for validating learned controller gains in physical flight tests.
    }
    \label{fig:quadcopter}
    \vspace{-1em}
\end{figure}

In this work, we use the 12dof model, described in Section \ref{sec:QuadDyn}, as the \textit{source environment} to learn the controller gains and a realistic Simulink model and the Holybro X500 V2 quadcopter, as shown in Figure \ref{fig:quadcopter}, as the \textit{target environment}.
In particular, we use the 12dof model, where $m = 1.56$ $\rm kg$ and the inertia is $J = {\rm diag} (0.03, 0.03, 0.05)$ $\rm kg \ m^2$ to learn the controller gains. 
These inertial properties are similar to the X500 Holybro quadcopter.
%

Recall that the control system consists of the outer and inner loops, as described in Section \ref{sec:autopilot}. 
The outer loop consists of three decoupled P-PI controllers, shown in Figure \ref{fig:cascaded-ppi}, to track the three-dimensional position references, and the inner loop consists of three decoupled P-PI controllers to track the three-dimensional Euler angle references.
Therefore, in total, 18 gains need to be learned. 
We use the CT-RCAC algorithm to learn the gains in a single trajectory, as discussed below.

\begin{remark}
\small
\it 
A 100-second simulation of the 12dof model takes approximately 3 seconds on an i7-13700HX Intel processor with 16 GB RAM, and thus the RCAC hyperparameters $G_\rmf(s),$ $R_z,$ and $P_0$ are tuned by a trivial grid-search method.  
The tuned RCAC hyperparameters used to learn the controller gains are given in Table \ref{tab:RCAC_para}.
\end{remark}

\begin{table}[h]
    \vspace{-1.8em}
    \caption{RCAC hyperparameters tuned with the 12 dof model in MATLAB.}
    \label{tab:RCAC_para}
     \vspace{0.5em}
    \centering
    \resizebox{0.8\columnwidth}{!}{%
    \begin{tabular}{|c|c|c|c|}
        \hline
        Hyperparameters &  $G_\rmf(s)$ & $R_z$ & $P(0)$ \\
        \hline
        & & & \\ [-1.75ex]
        Outer loop $(r_1, r_2)$ &  $\dfrac{1}{s+0.5}$ & $10^4$ & $10^3 I_{l_\theta}$ \\ [2ex]
        \hline
        & & & \\ [-1.75ex]
        Outer loop $(r_3)$ &  $\dfrac{1}{(s+1.5)(s+3)}$ & $10^{4}$ &  $10^5 I_{l_\theta}$ \\ [2ex]
        \hline
        & & & \\ [-1.75ex]
        Inner loop &  $\dfrac{1}{s+2}$ & $10^{4}$ & $10^3 I_{l_\theta}$ \\ [2ex]
        \hline
    \end{tabular}
    }
    \vspace{-1em}
\end{table}

\begin{figure}[h]
    \vspace{-0.5em}
    \centering
    \includegraphics[width=0.8\columnwidth]{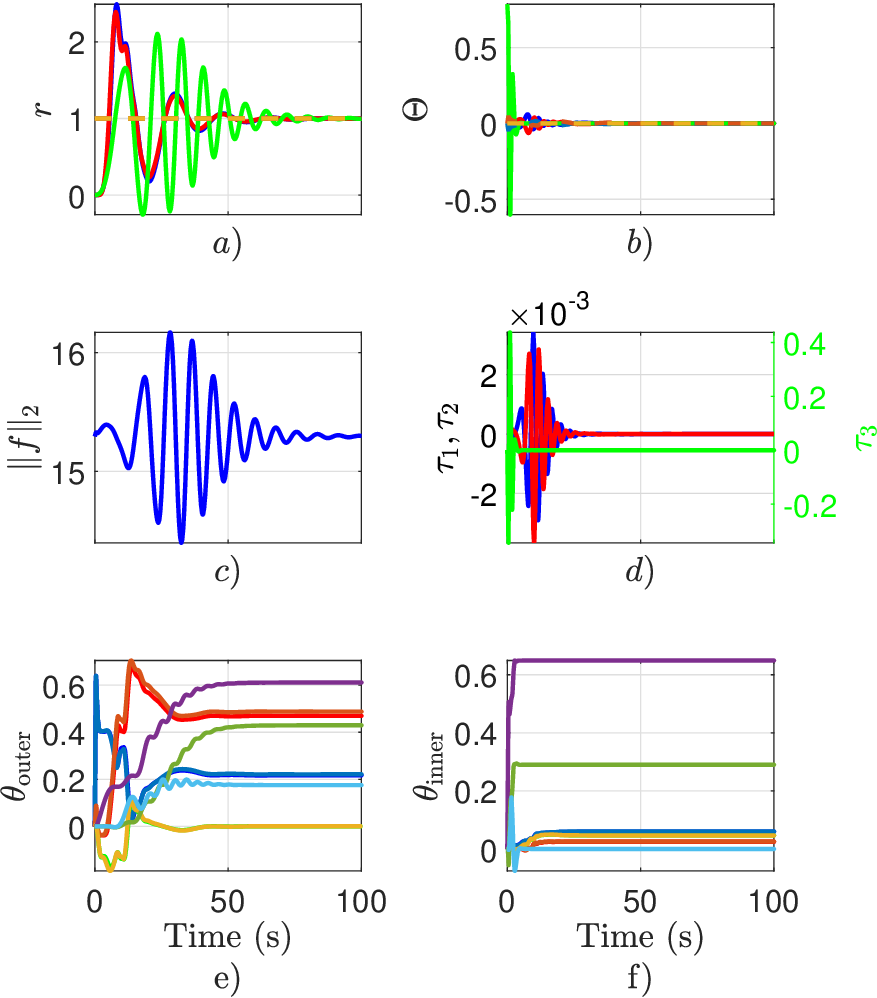}
    \caption{
    \textbf{Learning with a waypoint command}. 
    a) position response, 
    b) Euler angles response, 
    c) the magnitude of the force and  
    d) torque applied to the multirotor, respectively, 
    e) controller gains in the outer loop, and 
    f) controller gains in the inner loop.
    }
    \label{fig:quadcopter_simulation_step_position_data}
    \vspace{-1em}
\end{figure}

First, we consider a waypoint command, where the quadcopter is commanded to fly from an initial point to a final point. 
Note that a velocity command is not specified. 
In this work, we set the initial point to $(0,0,0)$ and the final point to $(1,1,1).$
Thus, the position command is, for all $t \geq 0,$ $r_{\rmd1} = r_{\rmd2} = r_{\rmd3} = 1.$
Figure \ref{fig:quadcopter_simulation_step_position_data} shows the a) position response, 
    b) Euler angles response, 
    c) the magnitude of the force and  
    d) torque applied to the multirotor, respectively, 
    e) controller gains in the outer loop, and 
    f) controller gains in the inner loop. 
The RCAC hyperparameters used in this case are shown in Table \ref{tab:RCAC_para}.
Note that all controller gains are initialized at zero.

\begin{figure}[h]
    \centering
    \includegraphics[width=0.8\columnwidth]{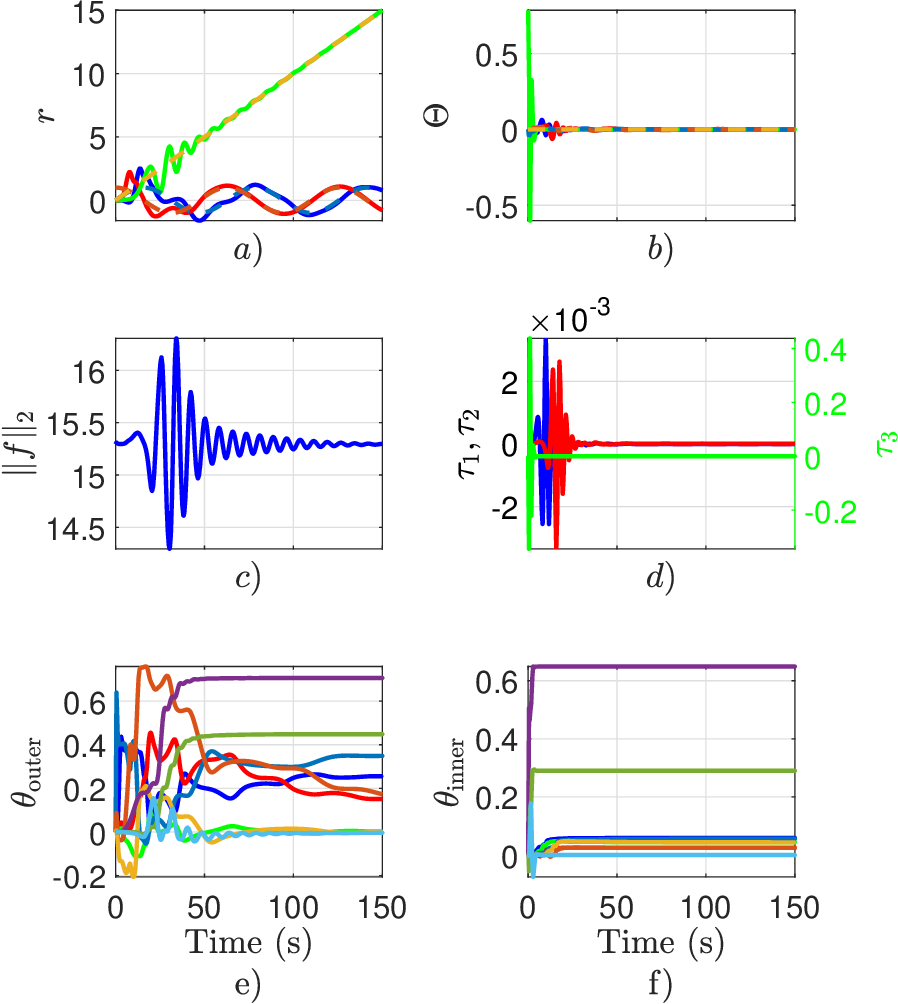}
    \vspace{0.5em}
    \caption{
    \textbf{Learning with a helical trajectory command}. 
    a) position response, 
    b) Euler angles response, 
    c) the magnitude of the force and  
    d) torque applied to the multirotor, respectively, 
    e) controller gains in the outer loop, and 
    f) controller gains in the inner loop. 
    }
    \vspace{-1em}
    \label{fig:quadcopter_simulation_traj_position_data}
\end{figure}

\begin{figure}[h]
    \centering
    \includegraphics[width=0.9\columnwidth]{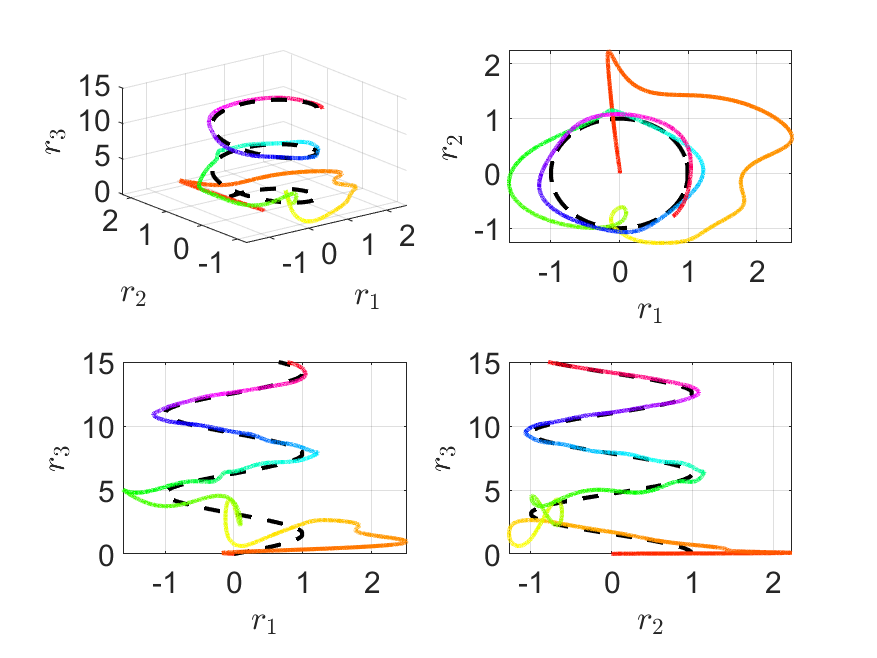}
    \caption{
    Trajectory-tracking response to the \textbf{helical trajectory command} with the learning controller. 
    }
    \label{fig:quadcopter_simulation_trajectory}
\end{figure}

Next, we consider a helical trajectory command, where the multirotor is commanded to follow a trajectory given by
$
    r_{\rmd1} = \cos (\omega t), 
    r_{\rmd2} = \sin (\omega t), 
$ and $
    r_{\rmd3} = \omega t.
$
In this work, we set $\omega = 0.1.$
Note that a velocity command is embedded in the position command. 
Figure \ref{fig:quadcopter_simulation_traj_position_data} shows the a) position response, 
    b) Euler angles response, 
    c) the magnitude of the force and  
    d) torque applied to the multirotor, respectively, 
    e) controller gains in the outer loop, and 
    f) controller gains in the inner loop. 
Note that the RCAC hyperparameters are not re-tuned.
Furthermore, to emphasize that no prior stabilizing controller is needed, all controller gains are initialized at zero.  
Figure \ref{fig:quadcopter_simulation_trajectory} shows the trajectory-tracking response to the {helical trajectory command} with the learning controller.
Note that the tracking performance improves the longer the simulation runs. 

The controller gains learned with the waypoint and the helical trajectory commands are shown in Table \ref{tab:converged_gains_helix}.

\begin{table}[h]
    \caption{Controller gains learned with the waypoint command and the helical trajectory command.}
    \label{tab:converged_gains_helix}
    \vspace{0.5em}
    \centering
    \begin{tabular}{|c|c|}
        \hline 
            Loop & Controller gains with waypoint command \\
        \hline
        &  \\ [-1.5ex]
        Outer   &  
        $\left\{
        \matl 
            0.2168 \\
            0.4697 \\
            0.0004
        \matr, 
        \matl 
            0.2212 \\
            0.4877 \\
            0.004
        \matr, 
        \matl 
            0.6114 \\
            0.4296 \\
            0.1753
        \matr
        \right\}$
        \\ [3.5ex]
        \hline
        &  \\ [-1.5ex]
        Inner  &  
        $\left\{
        \matl 
            0.0597 \\
            0.0249 \\
            0.0471
        \matr, 
        \matl 
            0.0607 \\
            0.0259 \\
            0.0464
        \matr, 
        \matl 
            0.6490 \\
            0.2902 \\
            1.025 \times 10^{-9}
        \matr
        \right\}$
        
        \\[3.5ex]

        \hline 
             & Controller gains with helical trajectory \\
        \hline
        &  \\ [-1.5ex]
        Outer  &  
        $\left\{
        \matl 
            0.2553 \\
            0.1514 \\
            0.0028
        \matr, 
        \matl 
            0.3489 \\
            0.1730 \\
            0.0007
        \matr, 
        \matl 
            0.7063 \\
            0.4485 \\
            0.0036
        \matr
        \right\}$
        \\ [3.5ex]
        \hline
        &  \\ [-1.5ex]
        Inner  &  
        $\left\{
        \matl 
            0.0587 \\
            0.0247 \\
            0.0446
        \matr, 
        \matl 
            0.0561 \\
            0.0247 \\
            0.0436
        \matr, 
        \matl 
            0.6490 \\
            0.2902 \\
            1.7 \times 10^{-8}
        \matr
        \right\}$
        \\ [3.5ex]
        \hline
    \end{tabular}
    \vspace{-1em}
\end{table}

\subsection{Simulation-based Verification of Learned Controller}
 
\begin{figure}[h]
    \centering
    \includegraphics[width=0.8\columnwidth]{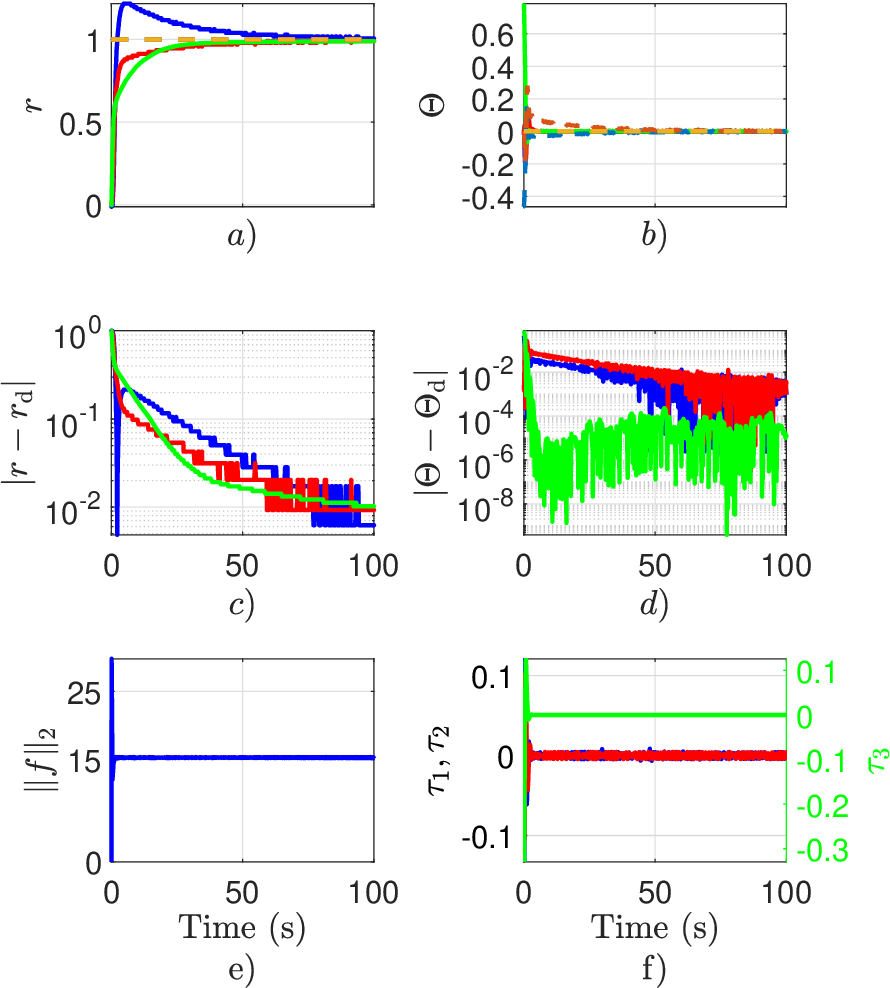}
    \vspace{0.5em}
    \caption{
    \textbf{Transferring learning to the Simulink model.}
    Waypoint tracking with the learned controller gains shown in Table 
    \ref{tab:converged_gains_helix}.
    a) position response, 
    b) Euler angles response, 
    c) the magnitude of the force and  
    d) torque applied to the multirotor, respectively.}
    \label{fig:quadcopter_px4_step_position_data}
    \vspace{-1em}
\end{figure}

\begin{figure}[h]
    \centering
    \includegraphics[width=0.8\columnwidth]{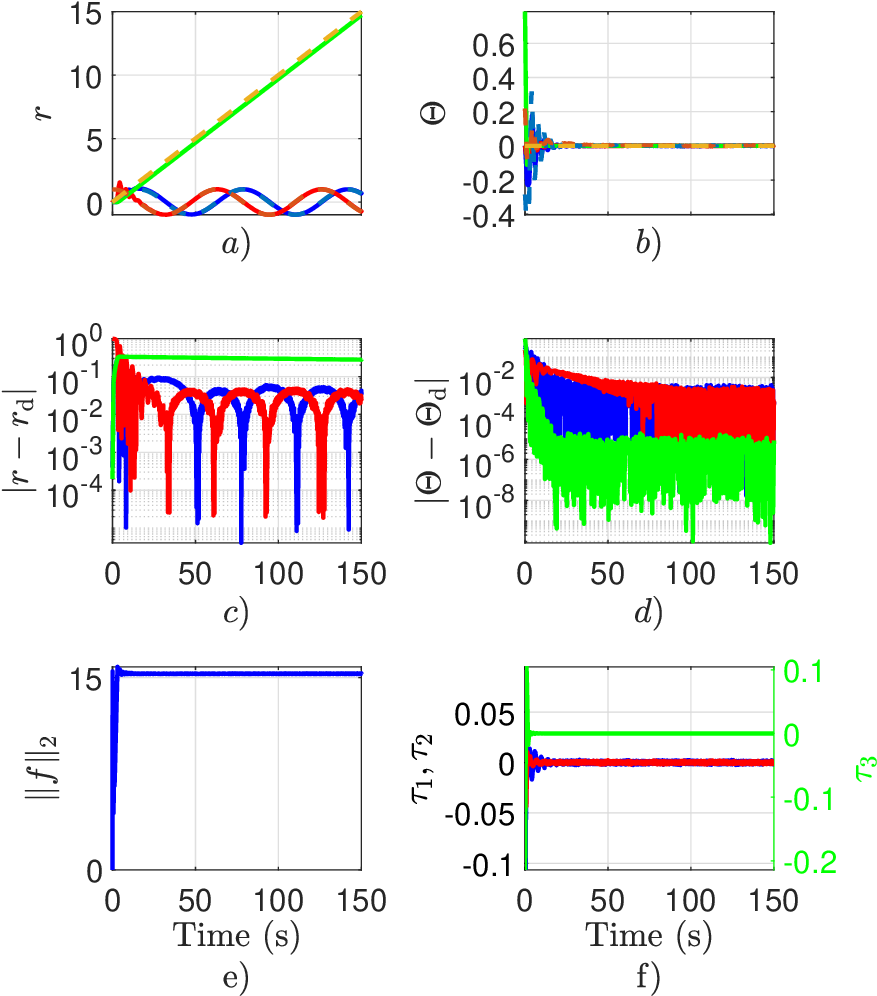}
    \vspace{0.5em}
    \caption{
    \textbf{Transferring learning to the Simulink model.}
    Helical trajectory tracking with the learned controller gains shown in Table \ref{tab:converged_gains_helix}. 
    a) position response, 
    b) Euler angles response, 
    c) the magnitude of the force and  
    d) torque applied to the multirotor, respectively.}
    \vspace{-1em}
    \label{fig:quadcopter_px4_traj_position_data}
\end{figure}

Next, we investigate the performance of the learned controller with a more realistic model by using the multirotor model implemented in Simulink \cite{matlabUAVmodel}. 
Note that the realistic Simulink model is thus the \textit{target environment} in the context of transfer learning. 
In addition to the multirotor's 12 dof nonlinear dynamics, the Simulink model considers realistic effects such as sensor noise, sensor delay, actuator dynamics, etc.
%
%
The default parameters modeling realistic sensor and actuator behaviors in the Simulink model are used.
Figures \ref{fig:quadcopter_px4_step_position_data} and \ref{fig:quadcopter_px4_traj_position_data} show the trajectory tracking response of the Simulink model to a waypoint and helical trajectory command, respectively, with the learned controller gains shown in Table \ref{tab:converged_gains_helix}. 
Note that the controller gains learned in the source environment yield a stable controller with acceptable transient performance. 
Additional tuning of the RCAC hyperparameters in the source environment may yield improved transient performance in the target environment. 

 \begin{figure}[h]
     \centering
     \includegraphics[width=0.9\columnwidth]{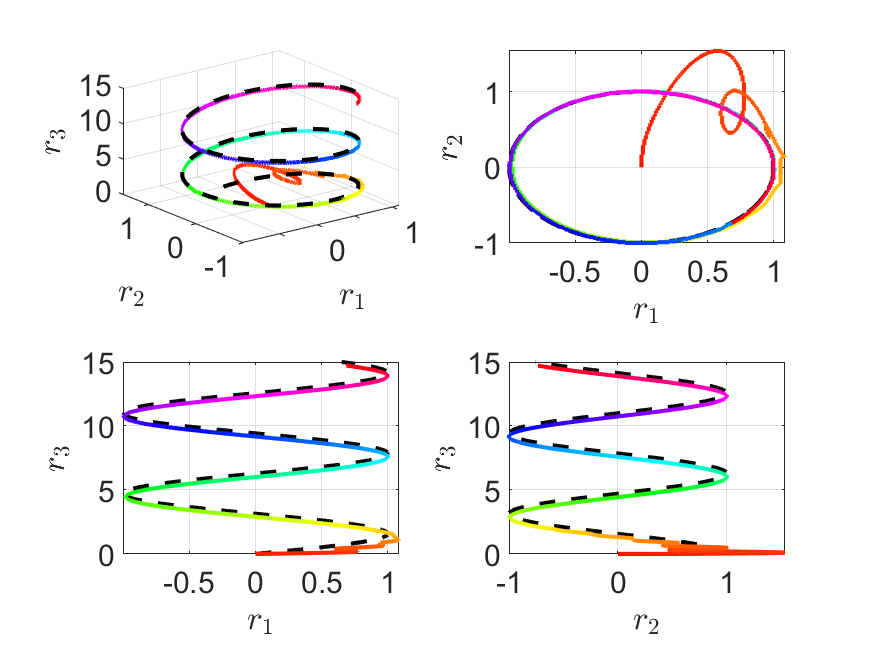}
     \caption{
     \textbf{Transferring learning to the Simulink model.}
     Trajectory-tracking response of the Simulink model with the learned controller.}
     \vspace{-2em}
     \label{fig:quadcopter_px4_trajectory}
 \end{figure}

\begin{figure}[h]
    \centering
    \includegraphics[width=0.9\columnwidth]{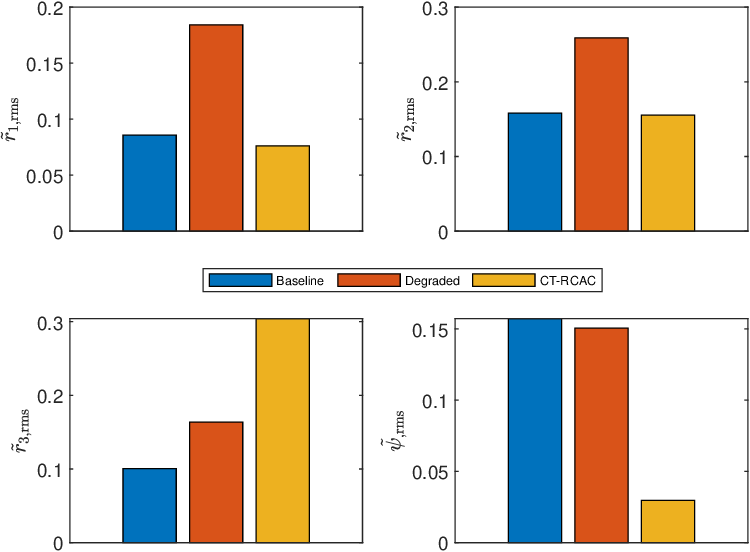}
    \vspace{0.5em}
    \caption{
    {RMS value of each component of the position error vector $\tilde r$ and the yaw error $\tilde \psi$ with the baseline, degraded, and the learned controller.}
    }
    \label{fig:position_error_compare}
    \vspace{-1em}
\end{figure}
\begin{figure}[h]
    \centering
    \includegraphics[width=0.9\columnwidth]{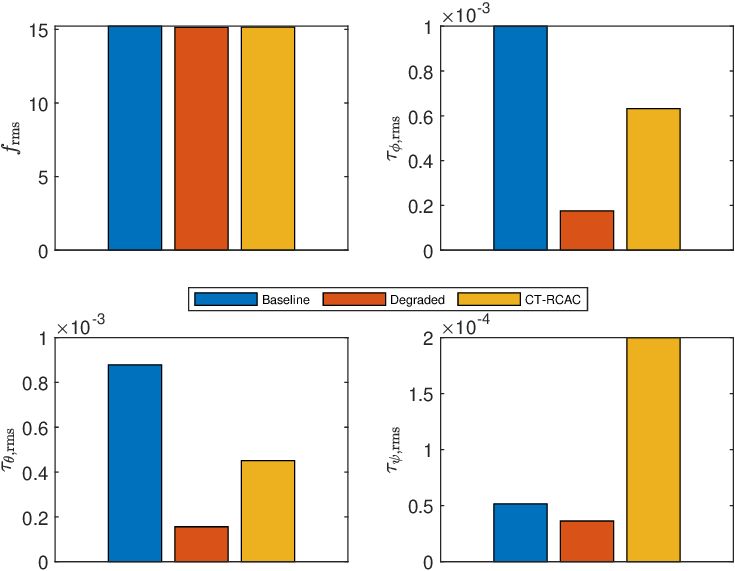}
    \vspace{0.5em}
    \caption{
    {RMS value of force $f$ and each component of the torque $\tau$ with the baseline, degraded, and the learned controller.}
    }
    \label{fig:input_compare}
    \vspace{-1em}
\end{figure}

Next, we compare the performance of the learned controller gains with the default controller gains in the PX4 simulation framework. 
The default PX4 gains are assumed to yield the baseline performance.
We also consider a degraded controller obtained by scaling all the default PX4 gains by a factor of 0.5.
Using the three controllers, the multirotor is commanded to follow the helical trajectory described above. 
The RMS value of each component of the position error vector $\tilde r$ and the yaw error $\tilde \psi$ is shown in Figure \ref{fig:position_error_compare}.
Note that the position error in the horizontal plane with the degraded controller increases by a factor of approximately two, while the learned controller restores the performance of the baseline controller, even though the learning process starts with zero prior gains. 
Similarly, the RMS value of force $f$ and each component of the torque $\tau$ is shown in Figure \ref{fig:input_compare}.

\subsection{Experimental Verification of Learned Controller}
\begin{figure}[h!]
    \centering
    \includegraphics[width=0.9\columnwidth]{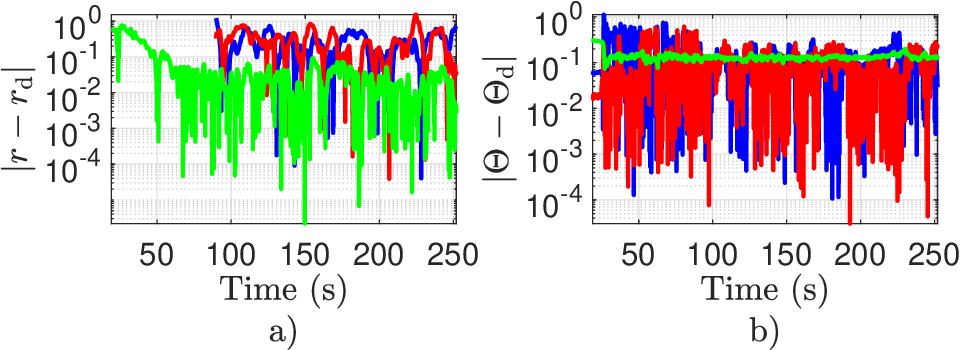}
    \caption{
    \textbf{Transferring learning to the X500 quadcopter.}
    Position and Euler angle error response with the X500 quadcopter to a waypoint command. 
    }
    \label{fig:experiment_flight_position_hold}
    \vspace{-1em}
\end{figure}

\begin{figure}[h!]
    \centering
    \includegraphics[width=0.9\columnwidth]{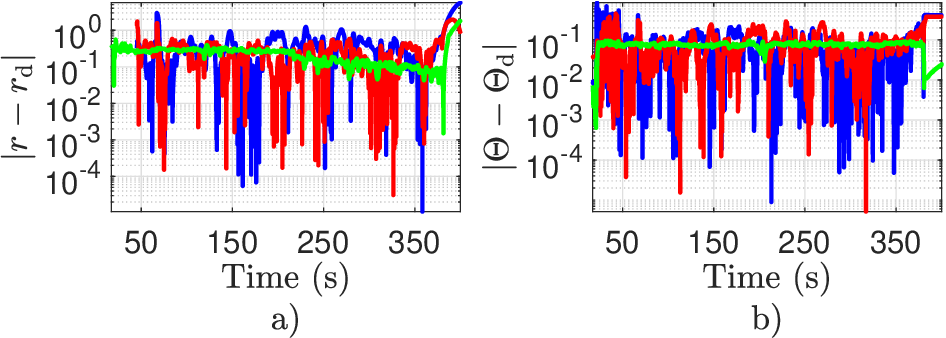}
    \caption{
    \textbf{Transferring learning to the X500 quadcopter.}
    Position and Euler angle error response with the X500 quadcopter to a helical trajectory command. 
    }
    \vspace{-2em}
    \label{fig:quadcopter_experiment_helix_errors}
\end{figure}

\begin{figure}[h]
    \centering
    \includegraphics[width=0.9\columnwidth]{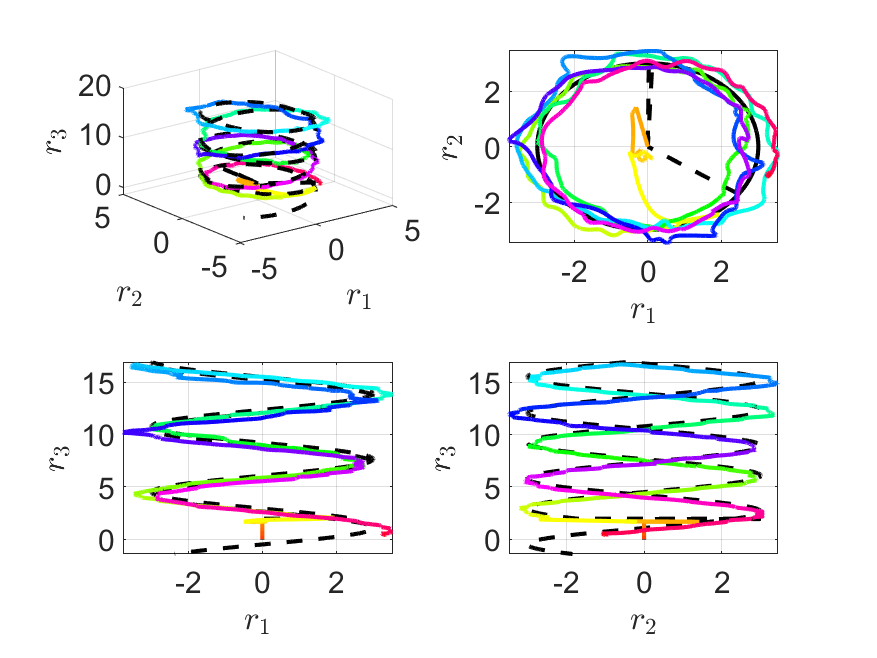}
    \caption{\textbf{Experimental flight with helical trajectory command.} Position response of the X500 quadcopter. 
    }
    \label{fig:quadcopter_experiment_helix_trajectory}
    \vspace{-1em}
\end{figure}

Finally, we experimentally evaluate the performance of the learned controller using the Holybro X500 V2 airframe. The flight tests are conducted outdoors under windy conditions near the UMBC campus. A Pixhawk 6C, operating at a sampling rate of 250 Hz, is utilized for onboard attitude and position estimation, as well as flight controller implementation.   
The physical quadcopter is thus the \textit{target environment} in the context of transfer learning. 
In this work, we use the Matlab UAV toolbox package for PX4 autopilot to implement the control architecture shown in Figure \ref{fig:cascaded-ppi} in each of the controllers in the inner and the outer loop.
The gains used in the nine controllers are shown in Table \ref{tab:converged_gains_helix}. 
Figures \ref{fig:experiment_flight_position_hold} and \ref{fig:quadcopter_experiment_helix_errors} show the position and Euler angle errors for the waypoint
\footnote{
\href{https://www.youtube.com/watch?v=djQX7IGlqbM}{https://www.youtube.com/watch?v=djQX7IGlqbM}
}
and helical trajectory
\footnote{
\href{https://www.youtube.com/watch?v=NDRqliiWJmY}{https://www.youtube.com/watch?v=NDRqliiWJmY}
}
commands, respectively. 
Figure \ref{fig:quadcopter_experiment_helix_trajectory} shows the trajectory tracking response of the quadcopter to the helical trajectory command. 
Note that the controller gains learned in the source environment yield a stable controller with acceptable transient performance. 
Additional tuning of the RCAC hyperparameters in the source environment may yield improved transient performance in the target environment. 

%

\section{Conclusions}
\label{sec:conclusions}

This paper presented the application of the continuous-time retrospective cost adaptive control technique 
%
to learn a set of stabilizing controller gains for an ideal 12dof multirotor model, and implemented these learned gains to contorl a realistic multirotor model within the Simulink UAV toolbox and the Holybro X500 V2 quadcopter with a Pixhawk 6C autopilot.
In the context of transfer learning, the ideal 12dof model served as the source environment, and the Simulink model and the X500 quadcopter served as the target environment.
To improve training performance, the inertial properties of the ideal 12dof model were set similar to those of the Holybro X500 V2 quadcopter. 
This paper showed that the controller gains learned in the source environment yielded a stable controller for the target environment with reasonable transient performance using a single learning experiment without the need to iterate between the source and target environment. 

Assessing the quality of the learned controller remains an open problem. 
Future work is thus focused on developing metrics to quantify the quality and performance of the learned controller as the target system is increasingly different from the source system.

\printbibliography

\end{document}

%% file: PackagesCommands.tex
\usepackage{amsmath} 
\usepackage{amsfonts}
\usepackage{amsthm}

\newtheorem{proposition}{Proposition}[section]
\theoremstyle{definition}

\newtheorem{remark}{Remark}

\usepackage{float} 
\usepackage[skip=0em,font=footnotesize]{caption}

\usepackage[margin = 1in]{geometry}

\usepackage{tikz}
\usetikzlibrary{shapes,arrows,calc,positioning}
\tikzstyle{bigblock} = [draw, fill=blue!20, rectangle, 
    minimum height=6em, minimum width=8em]
\tikzstyle{medblock} = [draw, fill=blue!20, rectangle, 
    minimum height=4em, minimum width=4em]    
\tikzstyle{mux} = [draw, fill=black!20, rectangle, 
    minimum height=5em, minimum width=0.1em]    
\tikzstyle{smallblock} = [draw, fill=blue!20, rectangle, 
    minimum height=2em, minimum width=3em]
    
\tikzstyle{data_block} = [draw, fill=green!20, rectangle, 
    minimum height=2em, minimum width=3em]
\tikzstyle{ops_block} = [draw, fill=blue!20, rectangle, 
    minimum height=2em, minimum width=3em]    
\tikzstyle{est_block} = [draw, fill=red!20, rectangle, 
    minimum height=2em, minimum width=3em]    
    
\tikzstyle{sum} = [draw, fill=blue!20, circle, node distance=1cm, minimum height=0.5cm]
\tikzstyle{signal} = [coordinate]
\tikzstyle{pinstyle} = [pin edge={to-,thin,black}]
\tikzstyle{block} = [draw, fill=blue!20, rectangle, 
    minimum height=3em, minimum width=9em]
\tikzstyle{blockS} = [draw, fill=blue!20, rectangle, 
    minimum height=3em, minimum width=4em]    
\tikzstyle{input} = [coordinate]
\tikzstyle{output} = [coordinate]
\usetikzlibrary{matrix}

\usetikzlibrary{positioning}
\usetikzlibrary{math}

\usepackage{hyperref}
\usepackage{xcolor}
\hypersetup{
    colorlinks,
    linkcolor={blue!100!black},
    citecolor={blue!50!black},
    urlcolor={blue!80!black}
}

\newcommand{\bc}{\begin{center}}
\newcommand{\ec}{\end{center}}
\newcommand{\benum}{\begin{enumerate}}
\newcommand{\eenum}{\end{enumerate}}
\newcommand{\nn}{\nonumber}
\newcommand{\matl}{\left[ \begin{array}}
\newcommand{\matr}{\end{array} \right]}

\renewcommand{\matl}{\begin{bmatrix}}
\renewcommand{\matr}{\end{bmatrix}}

\newcommand{\matls}{\left[ \begin{smallmatrix}}
\newcommand{\matrs}{\end{smallmatrix} \right]}
\newcommand{\isdef}{\stackrel{\triangle}{=}}

\newcommand{\rmT}{{\rm T}}

\newcommand{\rmd}{{\rm d}}

\newcommand{\rmf}{{\rm f}}

\newcommand{\rmi}{{\rm i}}

\newcommand{\rmp}{{\rm p}}

\newcommand{\BBR}{{\mathbb R}}

\newcommand{\SO}{{\mathcal O}}

\usepackage{pifont}
